\documentclass{webofc}
\usepackage[varg]{txfonts}   
\begin{document}
\title{Nature of excited $\Xi$ baryons with threshold effects}
%
%

\author{\firstname{Takuma} \lastname{Nishibuchi}\inst{1}\fnsep\thanks{\email{nishibuchi-takuma@ed.tmu.ac.jp}} \and
        \firstname{Tetsuo} \lastname{Hyodo}\inst{1}\fnsep\thanks{\email{hyodo@tmu.ac.jp}} 
}

\institute{Department of Physics, Tokyo Metropolitan University, Hachioji 192-0397, Japan
          }

\abstract{%
Spectroscopy of excited baryons with strangeness $S=-2$ is stimulated by recent experimental developments. Here we focus on the $\Xi(1620)$ which locates close to the $\bar{K}\Lambda$ threshold. To take into account the threshold effects, we construct the coupled-channels meson-baryon scattering amplitude where the $\Xi(1620)$ appears as a resonance. We demonstrate that the $\bar{K}\Lambda$ threshold effects distort the peak of the $\Xi(1620)$ resonance from the simple Breit-Wigner distribution.
}
\maketitle
\section{Introduction}

Recent progress in hadron spectroscopy enables us to find new resonances in the strange baryon sector. In fact, $\Lambda(1380)$, $\Lambda(2070)$, $\Lambda(2080)$ $\Lambda(2085)$, $\Sigma(2010)$, $\Omega(2012)$ are newly added in the listings by the Particle Data Group~\cite{Workman:2022ynf}. At the same time, theoretical frameworks have been elaborated to describe the excited hadrons as resonances in the scattering of hadrons~\cite{Hyodo:2020czb}. Thus, we are now in a position to study the strange baryon resonances systematically based on the experimental data. 

In the strangeness $S=-2$ sector, the Belle collaboration reported a new precise measurement of the $\pi\Xi$ spectrum in the $\Xi_c\rightarrow\pi\pi\Xi$ decay~\cite{Belle:2018lws}. The obtained spectrum shows clear peaks of the $\Xi(1620)$ and $\Xi(1690)$ resonances. In Ref.~\cite{Belle:2018lws}, the mass and width of the $\Xi(1620)$ are found to be $M_{R}=1610.4\pm 6.0^{+6.1}_{-4.2}$ MeV and $\Gamma_{R}=59.9\pm 4.8^{+2.8}_{-7.1}$ MeV, respectively. The properties of the neutral $\Xi(1690)$ was determined by the WA89 collaboration~\cite{Adamovich:1997ud} as $M_{R}=1686\pm 4$ MeV and $\Gamma_{R}=10\pm 6$ MeV which is adopted by the current PDG~\cite{Workman:2022ynf}. We however note that the peaks of the $\Xi(1620)$ and $\Xi(1690)$ appear just on top of the $\bar{K}\Lambda$ threshold ($\sim 1611$ MeV) and $\bar{K}\Sigma$ threshold ($\sim 1689$ MeV), respectively. It is known that the spectrum of a near-threshold resonance can be distorted by various kinematical effects at the threshold~\cite{Guo:2019twa}. Thus, the properties of these $\Xi$ resonances should be determined with proper treatment of the meson-baryon thresholds. 

In this work, we examine the spectrum of the $\Xi(1620)$ resonance by employing the chiral unitary approach~\cite{Kaiser:1995eg,Oset:1998it,Oller:2000fj,Hyodo:2011ur}, which describes the coupled-channels meson-baryon scattering including the threshold effects. Based on the previous work on the $\Xi(1620)$ in Ref.~\cite{Ramos:2002xh}, we construct a model which has a resonance pole of the $\Xi(1620)$ in the energy region reported in Ref.~\cite{Belle:2018lws}. Through the comparison with the Breit-Wigner distribution, we discuss the effect of the $\bar{K}\Lambda$ threshold for the spectrum of the $\Xi(1620)$.

\section{Formulation}

In the chiral unitary approach~\cite{Kaiser:1995eg,Oset:1998it,Oller:2000fj,Ramos:2002xh}, the coupled-channels $s$-wave meson-baryon scattering amplitude $T_{ij}(W)$ at total energy $W$ is given by the solution of the scattering equation
\begin{align}
T_{ij}(W)=V_{ij}(W)+\sum_{k}^{}V_{ik}(W)G_{k}(W,a_{k})T_{kj}(W), 
\end{align}
where the indices $i,j,k$ represent the meson-baryon channels. For the $\Xi(1620)$ with the strangeness $S=-2$ and isospin $I=1/2$, there are four channels in the isospin basis, $\pi\Xi,\bar{K}\Lambda,\bar{K}\Sigma$, and $\eta\Xi$. For the interaction kernel $V_{ij}(W)$, we use the Weinberg-Tomozawa interaction which satisfies the chiral low-energy theorem. The divergence of the loop function $G_i(W,a_i)$ is tamed by the dimensional regularization, and the finite part is specified by the subtraction constant $a_{i}$ in each channel.

Because the structure of the Weinberg-Tomozawa interaction is model-independently determined by chiral symmetry, four subtraction constants $a_{i}$ in the loop function are the free parameters in this model. In Ref.~\cite{Ramos:2002xh}, the subtraction constants are determined as $a_{i}=-2$ in all channels at the regularization scale $\mu=630$ MeV, based on the natural size argument in Ref.~\cite{Oller:2000fj}. The pole of the $\Xi(1620)$ is then found at $W=1607-140i\ {\rm{MeV}}$, which corresponds to $M_{R}=1607$ MeV and $\Gamma_{R}=280$ MeV. In view of the new Belle result~\cite{Belle:2018lws},  which indicates the narrower decay width, it is necessary to update the model. 

\section{Numerical results}

\begin{figure}[tbp]
    \centering
    \includegraphics[width=100mm,bb=0 0 842 595]{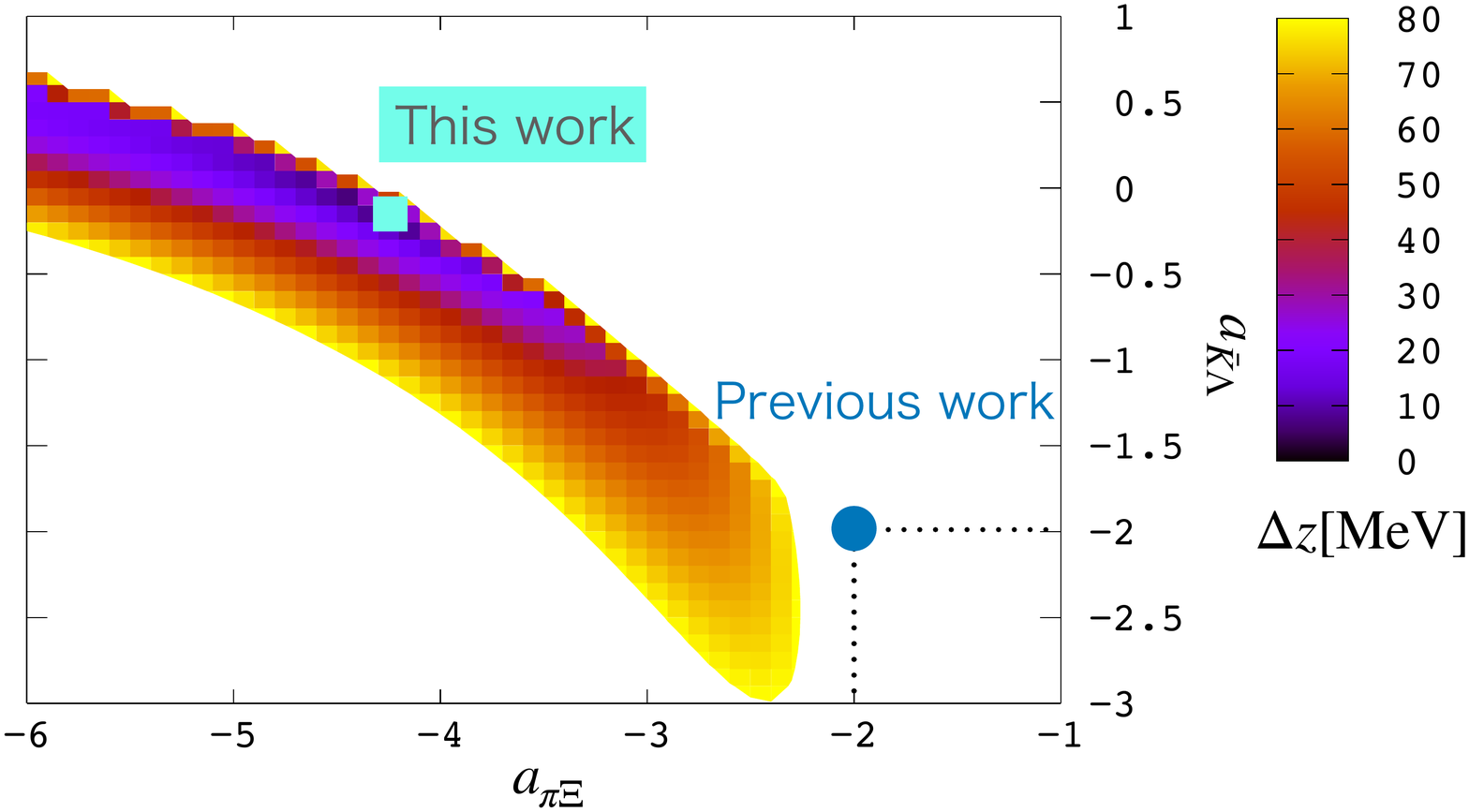}
 \vspace{-7mm}
    \caption{The density plot of the deviation of the pole position $\Delta z$ in the $a_{\pi\Xi}$-$a_{\bar{K}\Lambda}$ plane.}
  \label{21}
\end{figure}

Here we aim at constructing a model which is consistent with the Belle data~\cite{Belle:2018lws}. As a first trial, we assume that the pole of the $\Xi(1620)$ locates at $z_{\rm{ex}}=1610-30i$ MeV motivated by the Belle result, and try to reproduce $z_{\rm{ex}}$ in the chiral unitary approach. The pole position of the scattering amplitude of chiral unitary approach $z_{\rm th}$ depends on the subtraction constants $a_{i}$. Starting from the model of Ref.~\cite{Ramos:2002xh}, we vary the subtraction constants in the $\pi\Xi$ and $\bar{K}\Lambda$ channels, to which the pole position $z_{\rm th}$ is sensitive. To quantify the deviation of the pole positions, we define $\Delta z$ as
\begin{align}
\Delta z=|z_{\rm th}-z_{\rm{\rm ex}}|.
\label{eq:Deltaz}
\end{align}
In Fig.~\ref{21}, we show the density plot of $\Delta z$ in the $a_{\pi\Xi}$-$a_{\bar{K}\Lambda}$ plane. We find that $\Delta z$ is minimized by $(a_{\pi\Xi},a_{\bar{K}\Lambda})=(-4.19,-0.14)$, which gives $z_{\rm th}=1610-30i\ {\rm{MeV}}$ in agreement with $z_{\rm ex}$. 

\begin{figure}[tbp]
  \begin{minipage}[tbph]{0.5\linewidth}
    \centering
    \includegraphics[width=71mm,bb=0 0 842 595]{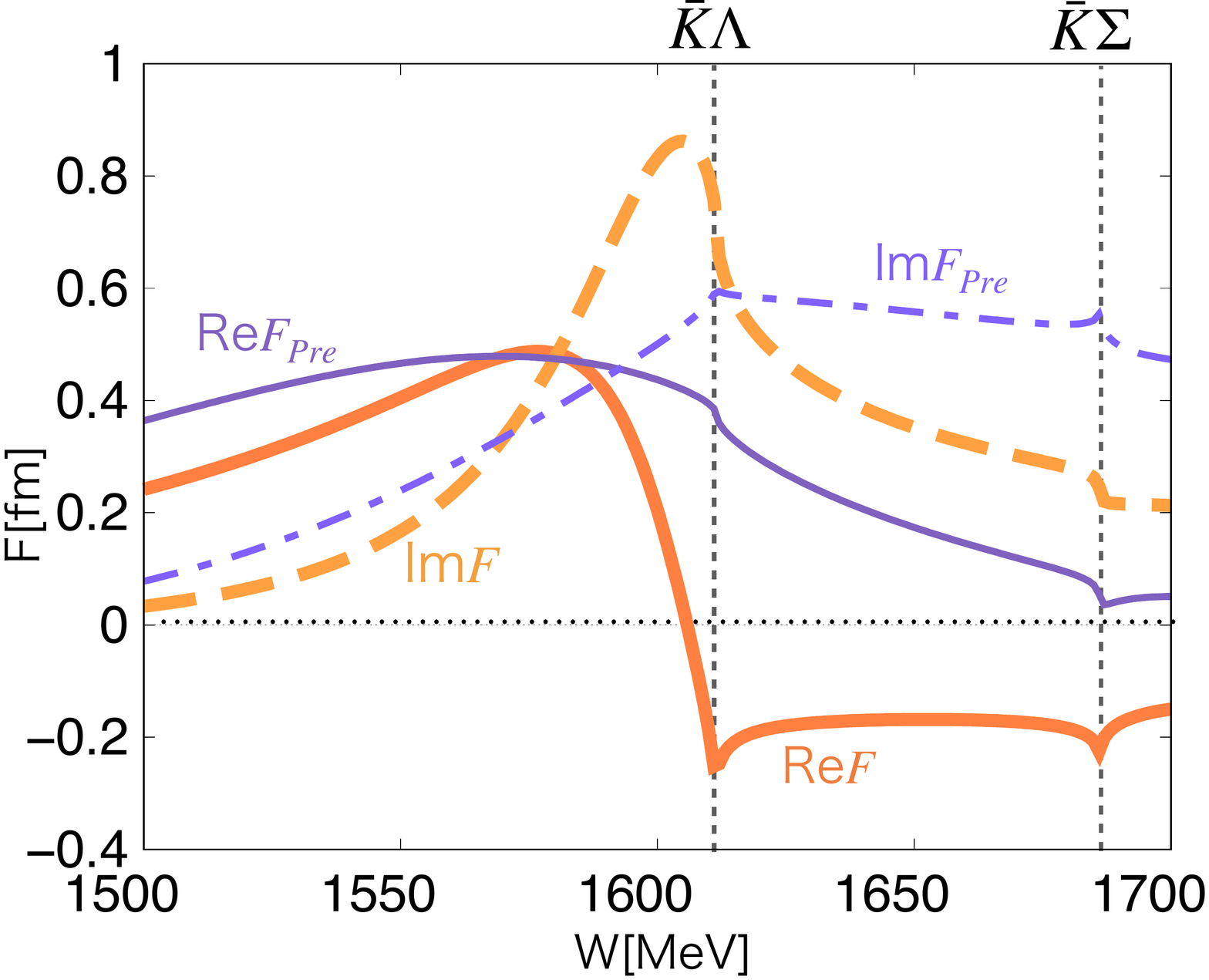}
  \end{minipage}
  \begin{minipage}[tbph]{0.5\linewidth}
    \centering
    \includegraphics[width=71mm,bb=0 0 842 595]{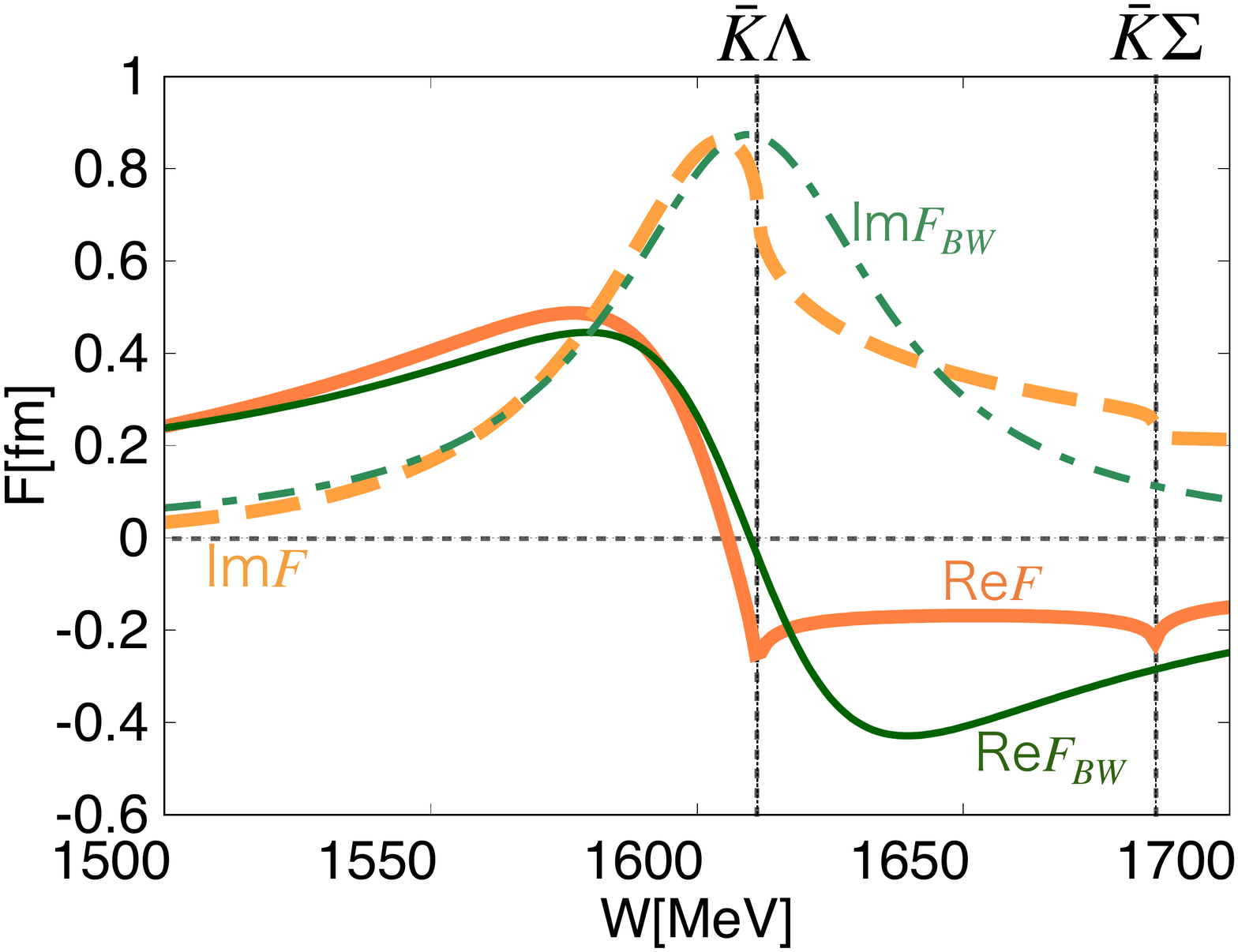}
  \end{minipage}
  \caption{Left: A comparison of the $\pi\Xi$ scattering amplitude in this work $F$ (thick lines) with the one from Ref.~\cite{Ramos:2002xh} $F_{Pre}$ (thin lines). The real (imaginary) part of the  scattering amplitude is denoted by the solid (dashed) lines. Vertical dotted lines indicate the $\bar{K}\Lambda$ and $\bar{K}\Sigma$ thresholds. 
  Right: A comparison with the Breit-Wigner amplitude $F_{BW}$. The legends are the same as in the left panel. }
  \label{31}
\end{figure}

Let us examine the behavior of the scattering amplitude on the real energy axis. In the left panel of Fig.~\ref{31}, we show the real and imaginary parts of the scattering amplitude $F$ in the diagonal $\pi\Xi$ channel as functions of the energy $W$, in comparison with the model of Ref.~\cite{Ramos:2002xh}. The amplitude of Ref.~\cite{Ramos:2002xh} (thin lines) does not exhibit a resonance behaviour (peak of the imaginary part and zero crossing of the real part) in this energy region, because of the broad decay width $\sim 280$ MeV. In contrast, the amplitude in this work shows a clear resonance behaviour around 1600 MeV. Thus, by reducing the decay width, the $\pi\Xi$ spectrum (Im $F$) can exhibit a resonance peak, as observed by the Belle collaboration.

As mentioned in the introduction, the peak of the $\Xi(1620)$ is very close to the $\bar{K}\Lambda$ threshold. To examine the threshold effect, we compare the the scattering amplitude of our model with the amplitude generated by the Breit-Wigner distribution,
\begin{align}
T^{\rm BW}(W)=\frac{Z}{W-M_R+i\Gamma_R/2} .
\end{align}
We set $M_{R}=1610$ MeV and $\Gamma_{R}=60$ MeV to have the same pole position as in our model. The residue $Z$ is adjusted to scale the magnitude of the spectrum. In the right panel of Fig.~\ref{31}, we present the comparison of our results with the Breit-Wigner amplitude. It can be seen from the figure that the peak of the imaginary part does not coincide with the Breit-Wigner distribution, even though the pole positions are the same. In fact, the peak of the imaginary part of our model is at $W\sim 1604$ MeV, which is shifted from the real part of the pole position ($W=1610$ MeV). In this way, we quantitatively show that the $\bar{K}\Lambda$ threshold effect distorts the spectrum of the $\Xi(1620)$ resonance. 

\section{Summary}

In this work, we study the spectrum of the $\Xi(1620)$ resonance by taking the nearby $\bar{K}\Lambda$ threshold into account. Within the framework of the chiral unitary approach, we construct a model of the $\Xi(1620)$ whose resonance pole appears in the energy region indicated by the recent Belle data. Through the comparison with the pure Breit-Wigner distribution, we find that the peak structure of the $\Xi(1620)$ is highly affected by the existence of the $\bar{K}\Lambda$ threshold. We conclude that the determination of the pole position of the $\Xi(1620)$ should be performed by properly taking into account the $\bar{K}\Lambda$ threshold.

As a future perspective, it will be interesting to study the origin of the $\Xi(1620)$ resonance in our model. In Ref.~\cite{Hyodo:2008xr}, it was shown that the change of the subtraction constants in the chiral unitary approach can be related to the inclusion of the genuine quark states. By applying the natural renormalization scheme, it is possible to extract the genuine quark component. A large deviation of the subtraction constants (Fig.~\ref{21}) may indicate the importance of the genuine component.
The $\bar{K}\Sigma$ threshold effect for the $\Xi(1690)$ resonance can also be studied in the same manner, by referring to the models in Refs.~\cite{Garcia-Recio:2003ejq,Sekihara:2015qqa} which reproduce the $\Xi(1690)$. For the $\Xi(1690)$, it should be noted that the central value of the $\Xi(1690)$ determined by the WA89 collaboration is just in between the $K^{-}\Sigma^{+}$ ($\sim 1683$ MeV) and $\bar{K}^{0}\Sigma^{0}$ ($\sim 1690$ MeV) thresholds. For the quantitative discussion of the $\Xi(1690)$, therefore, the isospin symmetry breaking effects should play an important role. 
To analyze the $\pi\Xi$ spectrum of the experimental data by the Belle collaboration, we need theoretical framework to calculate the $\Xi_{c}\to \pi\pi\Xi$ decay, as developed in Ref.~\cite{Miyahara:2016yyh}. These subjects will be investigated to understand the nature of the $\Xi$ resonances in future.

%

\begin{thebibliography}{13}

\bibitem{Workman:2022ynf}
R.L. Workman (Particle Data Group), PTEP \textbf{2022}, 083C01 (2022).

\bibitem{Hyodo:2020czb}
T.~Hyodo, M.~Niiyama, Prog. Part. Nucl. Phys. \textbf{120}, 103868 (2021).

\bibitem{Belle:2018lws}
M.~Sumihama et~al. (Belle), Phys. Rev. Lett. \textbf{122}, 072501 (2019).

\bibitem{Adamovich:1997ud}
M.I. Adamovich et~al. (WA89), Eur. Phys. J. C \textbf{5}, 621 (1998).

\bibitem{Guo:2019twa}
F.K. Guo, X.H. Liu, S.~Sakai, Prog. Part. Nucl. Phys. \textbf{112}, 103757
  (2020).

\bibitem{Kaiser:1995eg}
N.~Kaiser, P.B. Siegel, W.~Weise, Nucl. Phys. A \textbf{594}, 325 (1995).

\bibitem{Oset:1998it}
E.~Oset, A.~Ramos, Nucl. Phys. A \textbf{635}, 99 (1998).

\bibitem{Oller:2000fj}
J.A. Oller, U.G. Meissner, Phys. Lett. B \textbf{500}, 263 (2001).

\bibitem{Hyodo:2011ur}
T.~Hyodo, D.~Jido, Prog. Part. Nucl. Phys. \textbf{67}, 55 (2012).

\bibitem{Ramos:2002xh}
A.~Ramos, E.~Oset, C.~Bennhold, Phys. Rev. Lett. \textbf{89}, 252001 (2002).

\bibitem{Hyodo:2008xr}
T.~Hyodo, D.~Jido, A.~Hosaka, Phys. Rev. C \textbf{78}, 025203 (2008).

\bibitem{Garcia-Recio:2003ejq}
C.~Garcia-Recio, M.F.M. Lutz, J.~Nieves, Phys. Lett. B \textbf{582}, 49 (2004).
\bibitem{Sekihara:2015qqa}
T.~Sekihara, PTEP \textbf{2015}, 091D01 (2015).

\bibitem{Miyahara:2016yyh}
K.~Miyahara, T.~Hyodo, M.~Oka, J.~Nieves, E.~Oset, Phys. Rev. C \textbf{95},
  035212 (2017).

\end{thebibliography}
%
%

\end{document}